\documentclass{article}
\usepackage{graphicx}
\graphicspath{%
    {converted_graphics/}
    {C:/1KGLaaa/aatempfold/Papers/SM2-Growth-5/}
}
\begin{document}

\begin{center}
{\huge Toward a Comprehensive Model of}\vskip3pt

{\huge Snow Crystal Growth Dynamics:}\vskip6pt

{\LARGE 2. Structure Dependent Attachment Kinetics near -5 C}\vskip16pt

{\Large Kenneth G. Libbrecht}\vskip4pt

{\large Department of Physics, California Institute of Technology}\vskip-1pt

{\large Pasadena, California 91125}\vskip-1pt

\vskip18pt

\hrule\vskip1pt \hrule\vskip14pt
\end{center}

\textbf{Abstract.} We present experimental data demonstrating the presence
of structure-dependent attachment kinetics (SDAK) in ice crystal growth from
water vapor near -5 C. Specifically, we find that the nucleation barrier on
the basal edge of a thin-walled hollow columnar crystal is approximately ten
times smaller than the corresponding nucleation barrier on a large basal
facet. These observations support the hypothesis that SDAK effects play an
important role in determining the growth morphologies of atmospheric ice
crystals as a function of temperature.

\section{Introduction}

In \cite{sdak3, sdak1} we described a crystal growth instability that
enhances the development of thin edges, thereby promoting the formation of
plate-like or hollow columnar crystal morphologies. This instability arises
when diffusion-limited growth is coupled with structure-dependent attachment
kinetics (SDAK), specifically when the nucleation barrier on a facet surface
decreases substantially as the facet width approaches atomic dimensions. In 
\cite{sdak3} we also presented experimental data confirming the presence of
the SDAK instability in the growth of ice crystals from water vapor near -15
C. The data in \cite{sdak3} indicate that the SDAK instability plays a key
role in the formation of thin plate-like crystals at -15 C, which is a
well-known feature in the snow crystal morphology diagram \cite%
{libbrechtreview}.

In \cite{standardmodel} we further presented a new and comprehensive
physical model that begins to explain the overall structure of the
morphology diagram, in particular the observed changes in growth morphology
as a function of temperature. The SDAK instability plays a central role in
this model, in that it connects the intrinsic growth rates of faceted
surfaces to the observed morphological changes with temperature. This model
makes a strong prediction that SDAK effects should be observable at
temperatures near -5 C, in much the same way that they were observed near
-15 C in \cite{sdak3}.

In the present paper we describe an investigation of ice growth at a
temperature of -5.15 C, which is at the needle peak in the morphology
diagram \cite{needlepeak}. We find clear evidence for SDAK effects on the
basal facets, suggesting that the SDAK instability is largely responsible
for the formation of thin-walled hollow columnar crystals near this
temperature. These results support the model in \cite{standardmodel}, and
strongly support the hypothesis that SDAK effects play an important role in
determining the growth morphologies of atmospheric ice crystals.

\section{Intrinsic Growth Rates at -5.15 C}

Following \cite{standardmodel}, we define the \textit{intrinsic growth rates}
of the basal and prism surfaces as the growth rates of infinite, clean,
dislocation-free, faceted ice surfaces in near equilibrium with pure water
vapor at a fixed temperature. We parameterize the surface growth velocities
using $v=\alpha _{surf}v_{kin}\sigma _{surf},$ where $v$ is the
perpendicular growth velocity, $v_{kin}(T)$ is a temperature-dependent
\textquotedblleft kinetic\textquotedblright\ velocity derived from
statistical mechanics, and $\sigma _{surf}$ is the water vapor
supersaturation relative to ice at the growing surface. The intrinsic
attachment coefficient $\alpha _{surf}$ is parameterized using $\alpha
_{surf}(\sigma _{surf},T)=A\exp (-\sigma _{0}/\sigma _{surf})$, and
measurements of the parameters $A(T)$ and $\sigma _{0}(T)$ for the basal and
prism facets are given in \cite{prisms}, yielding $(A,\sigma
_{0})_{basal}=(1\pm 0.3,0.75\pm 0.1\%)$ and $(A,\sigma
_{0})_{prism}=(0.15\pm 0.05,0.17\pm 0.06\%)$ at $T=-5.15$ C.

Note that the $\alpha _{surf}(\sigma _{surf},T)$ on both facets are
determined by the detailed molecular dynamics occurring at the ice surface,
describing the various physical processes by which water vapor molecules
become incorporated into the crystalline lattice. The functional form above
is appropriate when the attachment kinetics are limited mainly by the
nucleation of molecular layers on the faceted surfaces, and the nucleation
parameter $\sigma _{0}$ derives from the step energy associated with these
molecular layers \cite{libbrechtreview, prisms}. The fact that the growth
measurements in \cite{prisms} are so well described by a nucleation-limited
model suggests the absence of significant dislocations on our test crystals,
and that the measurements in \cite{prisms} therefore provide a good
estimation of the intrinsic growth rates of the principal facets of ice.

One question that arose during our investigation was whether the intrinsic
growth rates depended on atmospheric pressure, particularly for the basal
facet. In other words, is $\alpha \left( \sigma _{surf}\right) $ on a basal
surface affected by the addition of clean air at a pressure of one bar?
Since air is chemically quite inert, we expect that its presence should have
little affect on the molecular dynamics affecting the attachment kinetics.
Therefore we expect that the measured $A(T)$ and $\sigma _{0}(T)$ should be
unaffected by the presence of the surrounding air. Nevertheless, we felt
that this expectation should be tested experimentally.

\begin{figure}[htb] 
  \centering
  \includegraphics[width=5.0in,keepaspectratio]{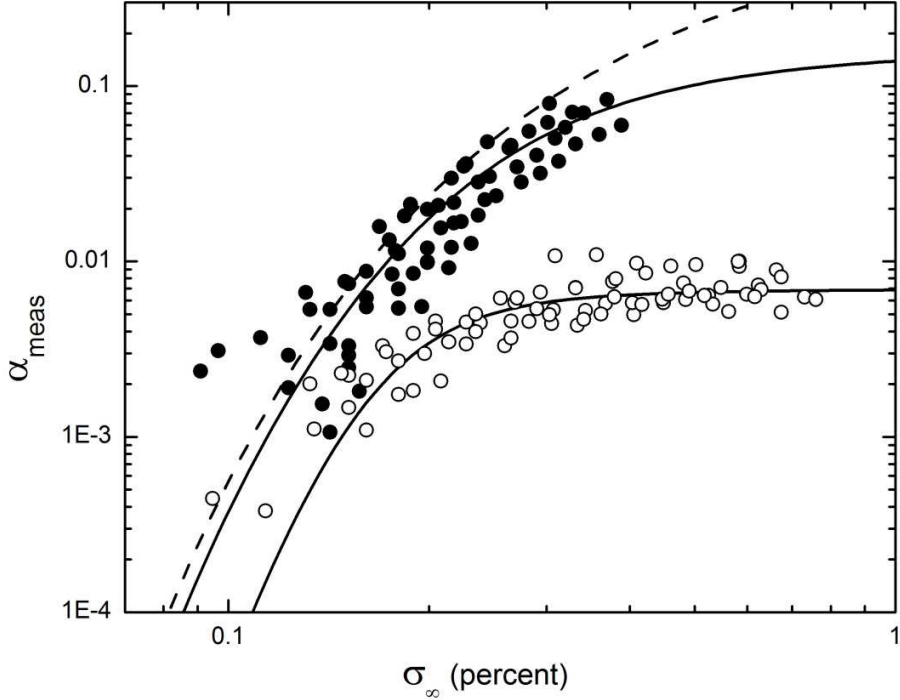}
  \caption{Measurements of the measured
attachment coefficient $\protect\alpha _{meas}\left( \protect\sigma _{\infty
}\right) =v/v_{kin}\protect\sigma _{\infty }$ for the basal facet at -5.15
C, as a function of supersaturation $\protect\sigma _{\infty }$ far from the
growing crystal. Solid circles show measurements made at a background
pressure of 0.03 bar, while open circles show measurements at 1 bar.
Theoretical curves are described in the text.}
  \label{basalintrinsic}
\end{figure}

Figure \ref{basalintrinsic} shows measurements of $\alpha _{meas}\left(
\sigma _{\infty }\right) =v/v_{kin}\sigma _{\infty }$ for the basal facet at
-5.15 C, following the notation in \cite{substinteract}, using the apparatus
described in \cite{apparatus}. Extracting $\alpha _{surf}(\sigma _{surf})$
from $\alpha _{meas}\left( \sigma _{\infty }\right) $ is complicated by the
fact that the observed crystal growth is limited by both the attachment
kinetics and by diffusion effects through the surrounding gas. At low
pressures the diffusion effects are relatively small, and can be removed
from the data as described in \cite{substinteract, prisms}. These data
analysis techniques are not adequate at pressures near one bar, however, so
we used an approximate diffusion modeling approach to compare with our data.

The dotted line in Figure \ref{basalintrinsic} shows $\alpha _{meas}\left(
\sigma _{\infty }\right) =\alpha _{surf}(\sigma _{surf})=A\exp (-\sigma
_{0}/\sigma _{\infty }),$ with $(A,\sigma _{0})_{basal}=(1,0.75\%),$ which
is the result from \cite{prisms}. This line represents the true intrinsic
growth rate of the basal facet at this temperature (within experimental
error). Put another way, taking $\alpha _{meas}\left( \sigma _{\infty
}\right) =\alpha _{surf}(\sigma _{surf})$ assumes no residual diffusion
effects, which is accurate in the limit of zero background pressure.

The nearby solid line in Figure \ref{basalintrinsic} shows%
\begin{equation}
\alpha _{meas}\left( \sigma _{\infty }\right) =\frac{\alpha _{surf}(\beta
\sigma _{surf})\alpha _{diff}}{\alpha _{surf}(\beta \sigma _{surf})+\alpha
_{diff}}  \label{model1}
\end{equation}%
which contains two diffusion correction factors. The $\alpha _{diff}$ term
corrects for the main diffusion effects, as described in \cite%
{substinteract, libbrechtreview}. For basal growth data as in Figure \ref%
{basalintrinsic}, the $\beta $ factor accounts for the fact that faster
growth of the nearby prism facets pulls down the supersaturation surrounding
the crystal, thus impeding the growth of the basal facets \cite%
{substinteract}. In Figure \ref{basalintrinsic} we used $\alpha _{diff}=0.2$
and $\beta =0.95,$ and these additional factors shift the dotted line to
better match the measured $\alpha _{meas}\left( \sigma _{\infty }\right) .$
Because these correction factors are quite small with a background pressure
of 0.03 bar, the model for $\alpha _{meas}\left( \sigma _{\infty }\right) $
(solid line) is fairly close to the intrinsic $\alpha _{surf}(\sigma
_{surf}).$

The lower solid line in Figure \ref{basalintrinsic} shows the same
functional form as in Equation \ref{model1}, but this time with values $%
\alpha _{diff}=0.007$ and $\beta =0.75.$ Using the known crystal sizes and
growth velocities to estimate the diffusion effects \cite{substinteract,
libbrechtreview} indicates that these fit values were reasonable for this
experiment. Since the same $\alpha _{surf}(\sigma _{surf})$ was used in the
model, this implies that $\alpha _{surf}(\sigma _{surf})$ for the
high-pressure data is indeed consistent with the $\alpha _{surf}(\sigma
_{surf})$ measured at lower pressure. Additional analysis, including
computer modeling of the crystal growth, allows us to place a limit of $%
\sigma _{0,basal} > 0.5$ percent from the high-pressure data in Figure %
\ref{basalintrinsic}.

Our overall conclusion from these data is that a background pressure of air
up to one bar seems to have little effect on the intrinsic basal $\alpha
_{surf}(\sigma _{surf}).$ A single $\alpha _{surf}(\sigma _{surf})$ can be
used to adequately model the measurements taken at low and high pressures,
as shown graphically in Figure \ref{basalintrinsic}. This experimental
conclusion agrees with our initial expectation that an inert background
should not change $\alpha _{surf}(\sigma _{surf})$ appreciably.
Nevertheless, we cannot completely rule out any pressure dependence in $%
\alpha _{surf}(\sigma _{surf}),$ since the diffusion effects present in the
data are very difficult to remove precisely. In addition, we cannot entirely
rule out residual chemical effects from impurities in the background gas,
although a separate investigation suggests that the impurity levels in our
experiments were too low to significantly affect the growth measurements 
\cite{impurities}.

\begin{figure}[htb] 
  \centering
  \includegraphics[width=2.9in,keepaspectratio]{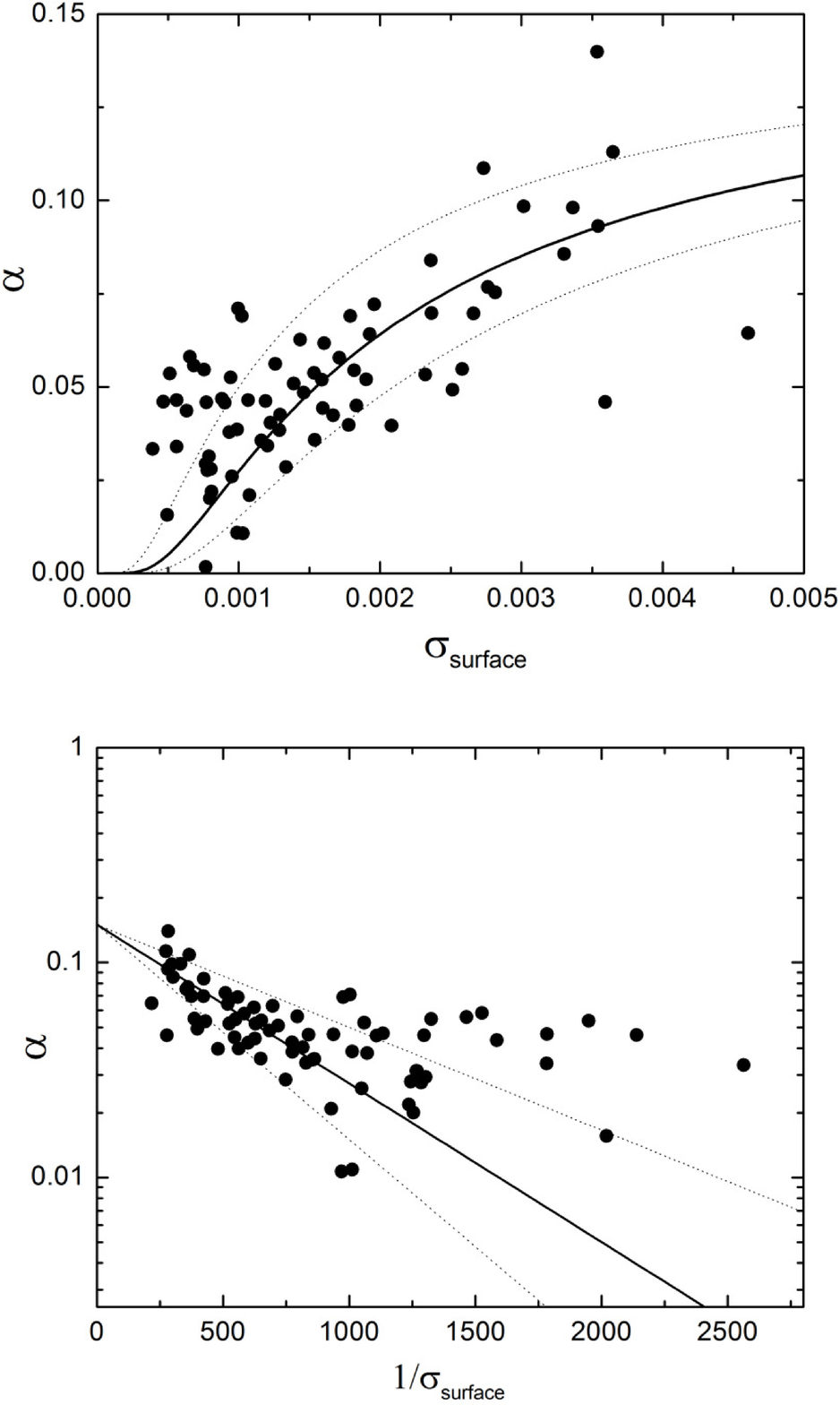}
  \caption{Measurements of $\protect\alpha %
_{surf}(\protect\sigma _{surf})$ for the prism facet at -5.15 C, from 
\protect\cite{prisms}. Lines show $(A,\protect\sigma _{0})=(0.15,0.17\pm
0.06\%).$ Growth of some crystals appeared to be anomolously high at low
supersaturations, so the low-$\protect\sigma $ points were given a somewhat
lower weight when fitting these data.}
  \label{intrinsicprism}
\end{figure}

Figure \ref{intrinsicprism} shows measurements of $\alpha _{surf}(\sigma
_{surf})$ for the prism facet from \cite{prisms}, along with curves showing $%
(A,\sigma _{0})_{prism}=(0.15,0.17\pm 0.06\%).$ How these data compare with
measurements at other temperatures is shown in \cite{prisms}. Additional
measurements at pressures near one bar (not shown) are also consistent with
our expectation that $(A,\sigma _{0})_{prism}$ is not substantially changed
by air background pressures up to one bar, again with the caveat that we
cannot positively exclude that there may be some pressure dependence in $%
\alpha _{surf}(\sigma _{surf}).$

From a combination of these and other supporting measurements from this
experiment \cite{prisms}, we therefore assume intrinsic growth rates
described by $(A,\sigma _{0})_{basal}=(1,0.75\%)$ and $(A,\sigma
_{0})_{prism}=(0.15,0.17\%)$ for the remainder of this investigation, and we
assume that these parameters are independent of background air pressure.

\section{SDAK Effects at -5.15 C}

To explore SDAK effects in growing ice crystals near -5 C, we again grew
small ice crystals on a sapphire substrate using the apparatus described in 
\cite{apparatus}, in air at a background pressure of one bar. Each run began
with an isolated, simple hexagonal prism crystal on the substrate, with one
prism facet resting on the substrate. Once the system was stable, the
supersaturation was increased and the subsequent growth was monitored, using
both direct imaging and laser interferometry \cite{prisms, apparatus}.

\begin{figure}[htb] 
  \centering
  \includegraphics[width=4.5in,keepaspectratio]{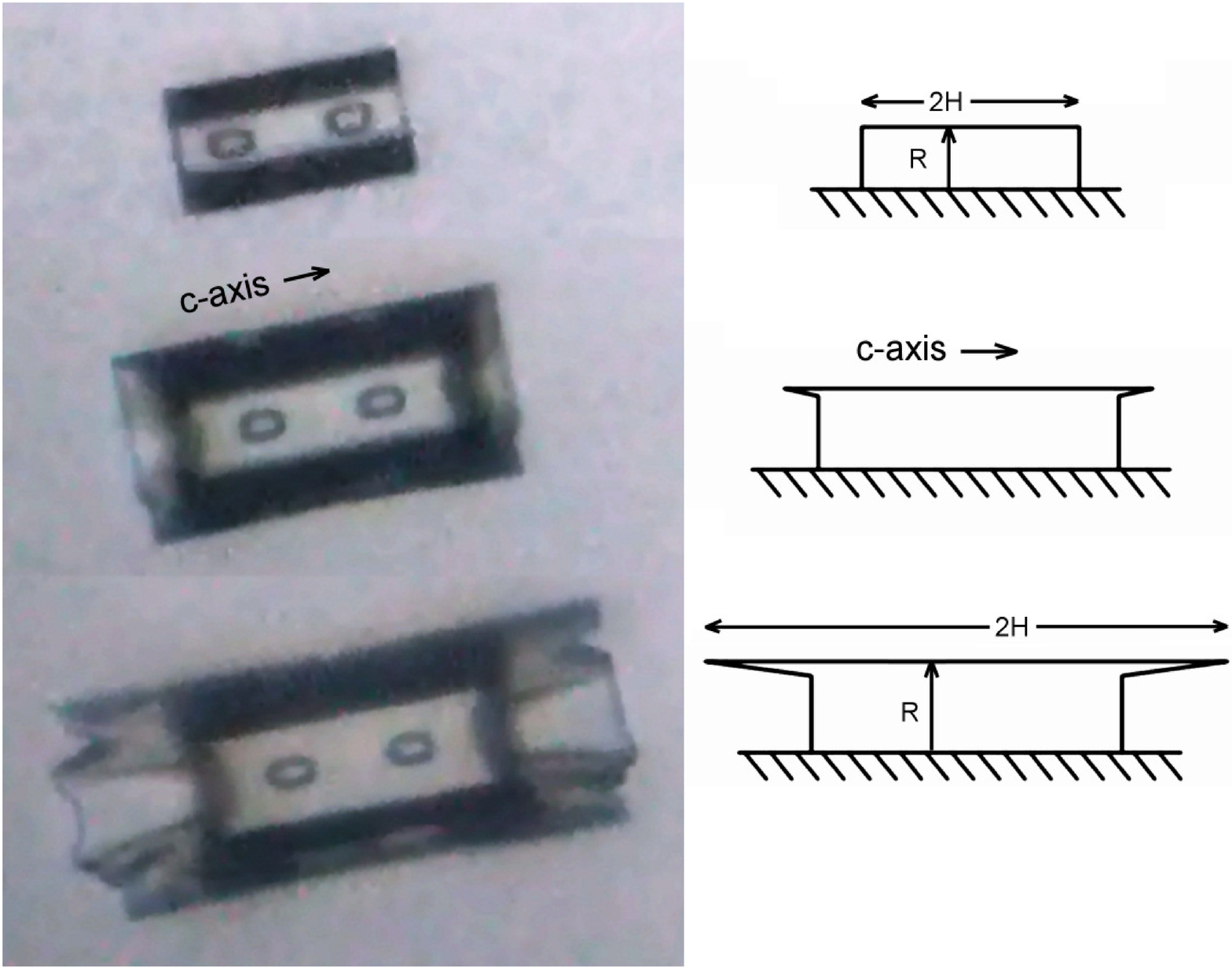}
  \caption{The three images on the left
show different stages of an ice crystal growing on a sapphire substrate. The
length of the prism along the c-axis in the top image is approximately 60 $%
\protect\mu $m, while the overall length of the structure in the lower image
is approximately 145 $\protect\mu $m. The elapsed time between the first and
last images was about four minutes. Two small enclosed bubbles in the
crystal are seen in all three images. The sketches on the left show
simplified cross-sections of the crystals, in a plane perpendicular to the
substrate. The effective radius $R$ and half-height $H$ of the crystal are
defined here. Note that $H$ is measured along the c-axis.}
  \label{sample}
\end{figure}

An example crystal from these measurements is shown in Figure \ref{sample}.
This crystal was initially grown at low supersaturation, and during this
phase the morphology remained that of a simple faceted prism. As the base
grew and filled in, the overall morphology became approximately that of half
a hexagonal prism crystal. The supersaturation was slowly increased with
time in this example, and above $\sigma _{\infty }\approx 1.5$ percent the
growth of the flat basal facets became unstable to basal hollowing. The
basal hollowing became quite pronounced as the supersaturation was
increased, yielding a thin-walled growth morphology, as seen in the lower
image in Figure \ref{sample}. To a reasonable approximation, the overall
morphology was essentially that of half a hollow column crystal, in this
case with quite thin columnar walls. The exact thickness of these walls,
along with other morphological details, could not be determined in this
experiment.

As with the experiments described in \cite{sdak3}, the initial sizes of our
test crystals were not constant, plus substrate interactions \cite%
{substinteract} varied somewhat from crystal to crystal. While the
qualitative growth behavior was quite similar for all crystals examined at a
given $\sigma _{\infty },$ the quantitative details depended on the initial
conditions in the experiment, which were different for each individual
crystal. For this reason, as in \cite{sdak3}, we found it most useful to
measure and model the growth of individual example crystals, rather than
measuring numerous crystals and forming averages. As described below, our
overall results did not depend on the specific crystals analyzed.

To model our growth measurements, we used the 2D cylindrically symmetric
cellular automata technique described in \cite{kglca}, again following the
procedures described in \cite{sdak3}. For each numerical model we input the
initial crystal size $(R_{0},H_{0}),$ the attachment coefficients $\alpha
_{prism}(\sigma _{surf})$ and $\alpha _{basal}(\sigma _{surf}),$ and a
constant supersaturation far from the crystal $\sigma _{\infty }.$ The
cellular automata technique then solved the diffusion equation around the
crystal and numerically \textquotedblleft grew\textquotedblright\ the
crystal, thus yielding the crystal size and morphology as a function of time.

\begin{figure}[tbp] 
  \centering
  \includegraphics[width=5.67in,height=5.25in,keepaspectratio]{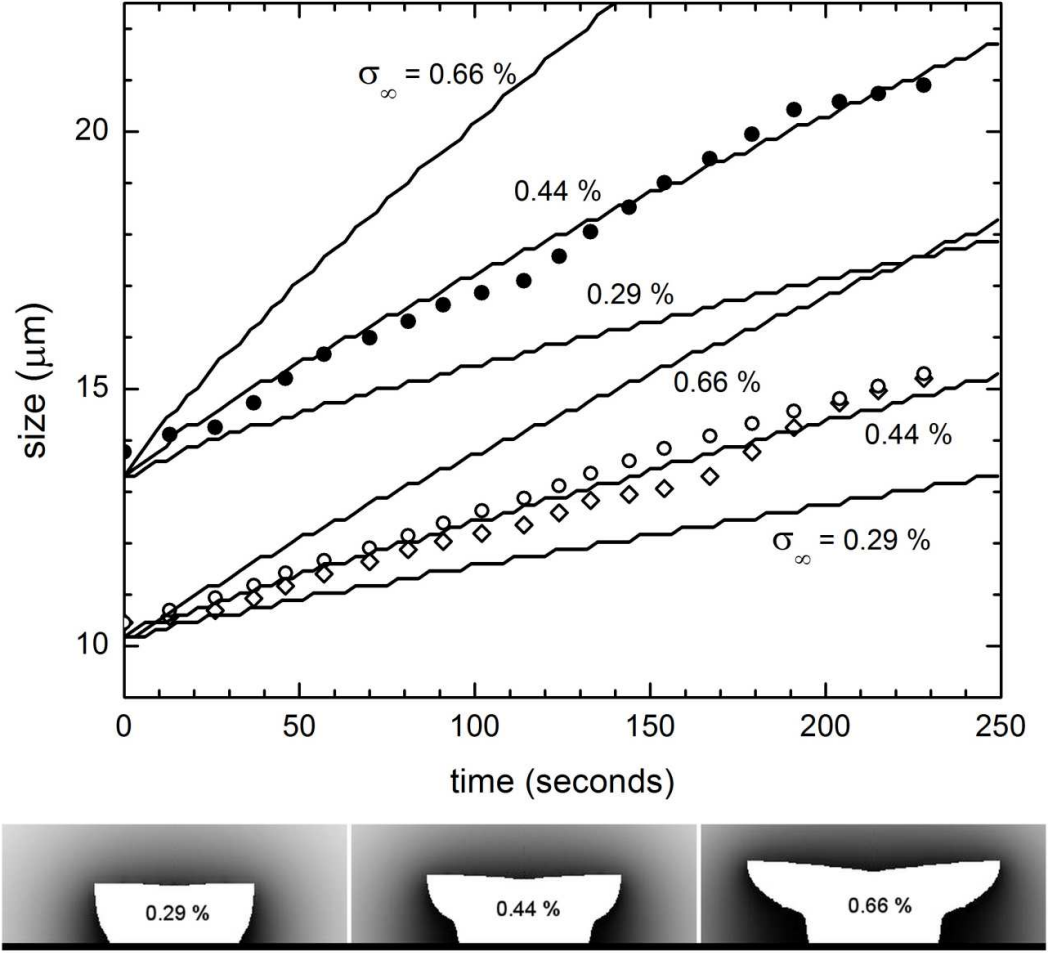}
  \caption{Data showing the growth of an ice
crystal on a substrate at a temperature of $T=-5.15$ C and a supersaturation
of $\protect\sigma _{\infty }=1.0$ percent, with a background air pressure
of 1 bar. The crystal morphology remained that of a simple faceted prism
throughout the run, similar to the top image in Figure \protect\ref{sample}.
The solid points show the measured half-length $H$ of the column as a
function of growth time. The open points show the effective radius $R$ of
the column, as measured by the distance between the substrate and the upper
prism facet (diamonds) and the observed half-width of the column along the
substrate (open circles). Lines show model crystal calculations using $(A,%
\protect\sigma _{0})_{basal}=(1,0.3\%),$ $(A,\protect\sigma %
_{0})_{prism}=(0.15,0.17\%),$ $(R,H)_{initial}=(10.2,13.3),$ and
supersaturations $\protect\sigma _{\infty }=$ 0.29, 0.44, and 0.66 percent,
as labeled. The images below the graph show calculated crystal cross
sections at $t=250$ for the different models, with the same orientation
shown in the sketches in Figure \protect\ref{sample}.}
  \label{xtal1sinf}
\end{figure}

\begin{figure}[tbp] 
  \centering
  \includegraphics[width=5.67in,height=5.25in,keepaspectratio]{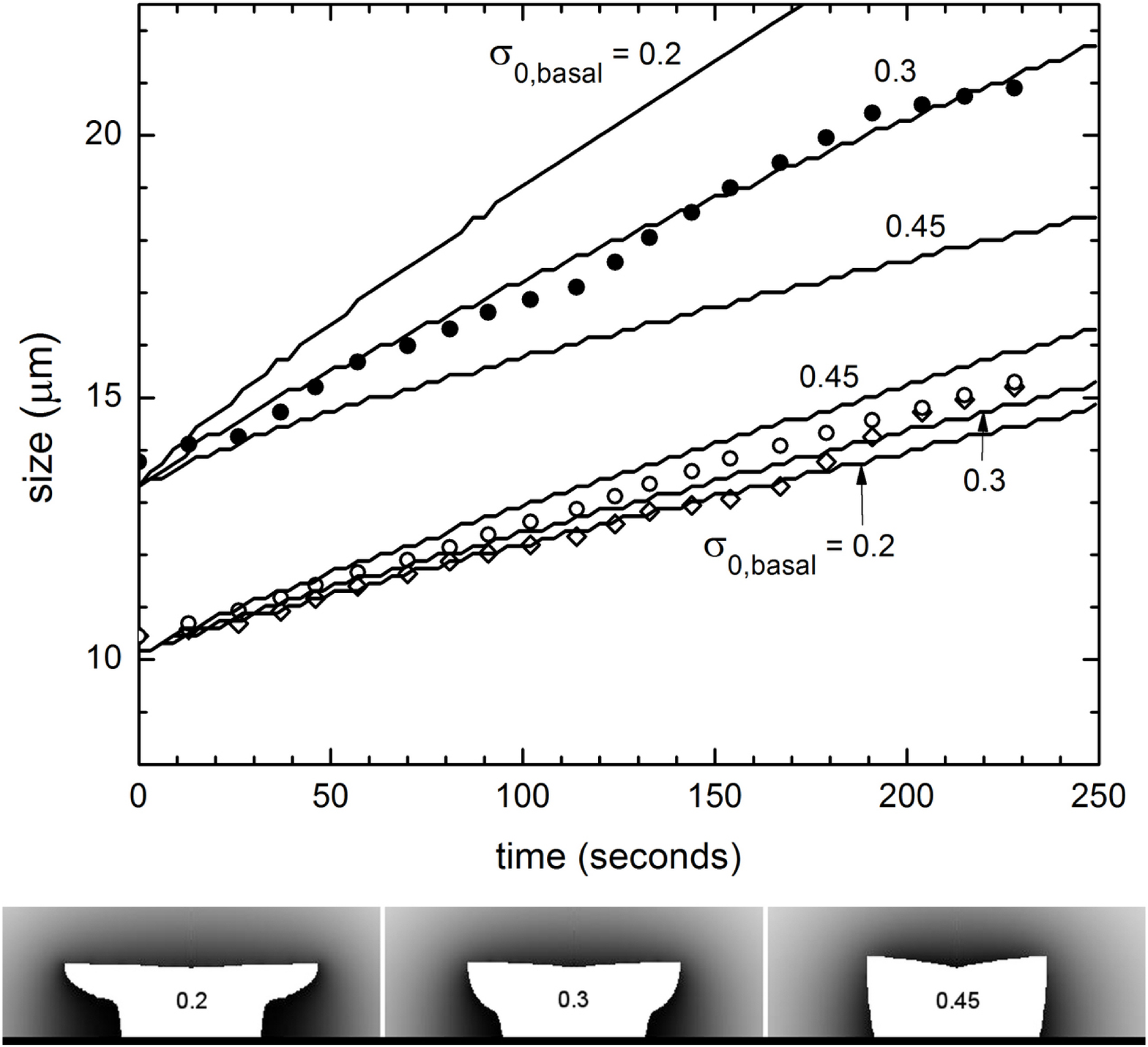}
  \caption{The same experimental data as in
Figure \protect\ref{xtal1sinf}, but plotted with different theoretical
models. Lines show model crystal calculations using $(A,\protect\sigma %
_{0})_{prism}=(0.15,0.17\%),$ $(R,H)_{initial}=(10.2,13.3),$ $A_{basal}=1,$ $%
\protect\sigma _{\infty }=$ 0.44 percent, with $\protect\sigma _{0,basal}=$
0.2, 0.3, and 0.45 percent, as labeled. The images below the graph again
show calculated crystal cross sections at $t=250$ for the different models.}
  \label{xtal1s0}
\end{figure}

Figures \ref{xtal1sinf} and \ref{xtal1s0} show measurements of a single test
crystal grown at $\sigma _{\infty }=1.0$ percent, where this value was
determined from experimental parameters \cite{apparatus}. The morphology of
this crystal remained essentially that of half a simple hexagonal prism
during the course of the experiment, similar to that shown in the top image
shown in Figure \ref{sample}. As the dimensions of the crystal increased
with time, we measured the effective prism radius $R$ and half-height $H$ as
defined in Figure \ref{sample}. The the radius was measured using two
methods -- from direct imaging of the half-width of the crystal, and by
using laser interferometry to measure changes in the distance between the
substrate and the top prism facet. From the latter measurements of $dR/dt,$
we used the initial $R_{0}$ from direct imaging and integrated $dR/dt$ to
produce an $R(t)$ from the interferometer data. As seen in the figures, the
two measurements of $R(t)$ gave similar results. Since substrate
interactions reduced the nucleation barrier on the side prism facets
somewhat \cite{substinteract}, we typically found that $R(t)$ measured from
direct imaging was slightly larger than $R(t)$ measured interferometrically.

Since the prism facets were large and flat on this crystal, in our models we
set $\alpha _{prism}$ to be that determined from the intrinsic growth
measurements described above, and we neglected any substrate interactions
for the prism facets in our modeling. The basal facets were also observed to
be quite flat in these crystals, so we would expect $\alpha _{basal}$ to be
close to the intrinsic value. However, the basal facets also intersected the
substrate, and it is certainly possible that substrate interactions reduced
the nucleation barrier on the basal facets, via the mechanism described in 
\cite{substinteract}. Indeed, since $\sigma _{0,basal}>\sigma _{0,prism},$
we expect that substrate interactions on the basal facet would perturb the
basal growth more than we observed on the prism facets.

In addition, we also know that the facets of a growing crystal are somewhat
convex, owing to diffusion effects. Because of this, the SDAK effect on the
nonflat basal facets could result in a smaller $\sigma _{0,basal}$ compared
to the intrinsic value. This possibility, along with possible substrate
interactions, means that $\sigma _{0,basal}$ could easily be smaller than
the intrinsic value. For this reason we kept $\sigma _{0,basal}$ as a model
variable in our calculations.

Figure \ref{xtal1sinf} shows three models in which we fixed $\alpha _{basal}$
and varied $\sigma _{\infty },$ centered about our best-fit model. Not
surprisingly, higher $\sigma _{\infty }$ values yielded faster growth rates
for both facets. Our best-fit model had $\sigma _{\infty }=0.44$ percent,
lower than the $\sigma _{\infty }$ = 1.0 percent set in the experiment. We
have come to understand this rough factor of two from previous experiments 
\cite{sdak3}. Most of the factor comes from the model itself, as various
effects tend to produce faster growth than seen experimentally. (For
example, one reason is that the outer boundaries of the model are fairly
close to the crystal, yielding faster growth rates than if the outer
boundaries were as far away as in the experiments.) To account for this
systematic modeling effect, along with uncertainties in the experimental $%
\sigma _{\infty },$ we adjusted the model $\sigma _{\infty }$ to fit the
data, thus yielding the best fit value $\sigma _{\infty }=0.44$ percent.

Figure \ref{xtal1s0} shows models in which we fixed $\sigma _{\infty }$ and
varied $\sigma _{0,basal},$ again centered about our best-fit model. In this
figure we see that reducing the nucleation barrier on the basal facet
increased the basal growth, as expected. The increased basal growth then
robbed water vapor from the neighboring prism facets, reducing their growth,
also as we would expect. By adjusting both $\sigma _{\infty }$ and $\sigma
_{0,basal},$ we produced our best-fit model with $\sigma _{\infty }=0.44$
percent and $\sigma _{0,basal}=0.3$ percent.

Our overall conclusion with this crystal is that it is reasonably well fit
using the intrinsic growth rates, with some caveats. The model $\sigma
_{\infty }$ is about a factor of two lower than we set in our experiment,
and we understand this factor as arising from modeling systematics along
with possible experimental systematics, as we described in \cite{sdak3}. In
addition, our fit $\sigma _{0,basal}$ was about a factor of two lower than
the intrinsic value. This reduced nucleation barrier most likely resulted
from substrate interactions with the basal facet \cite{substinteract}, which
we could not control in this experiment. Thus while our experimental and
modeling systematics are not negligible, the growth of this crystal is
generally consistent with expectations based on the measured intrinsic facet
growth rates, for both the prism and basal facets. In particular, no SDAK
effects, or perhaps only small SDAK effects, are needed to explain the
growth of this low-$\sigma _{\infty }$ crystal.

\begin{figure}[tbp] 
  \centering
  \includegraphics[width=5.67in,height=5.27in,keepaspectratio]{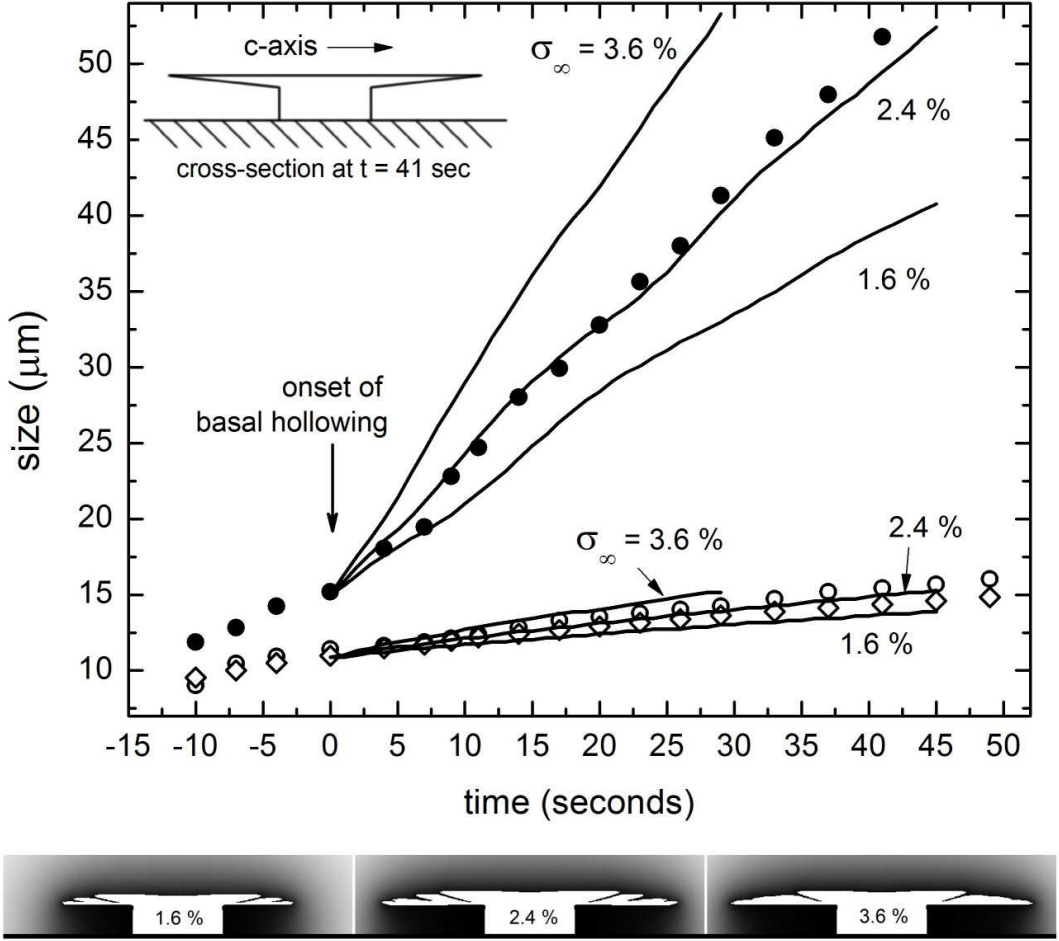}
  \caption{Data and models showing the
growth of an ice crystal on a substrate at a temperature of $T=-5.15$ C and
a supersaturation of $\protect\sigma _{\infty }=3.9$ percent, with a
background air pressure of 1 bar. The crystal was initially a simple
hexagonal prism, and it subsequently grew into a partial hollow column
morphology with thin columnar walls, similar to the example shown in Figure 
\protect\ref{sample}. The time axis was shifted so the onset of basal
hollowing occurred at $t=0.$ The solid points show the half-length $H$ of
the column as a function of growth time. The open points show the effective
radius $R$ of the column, as measured by the distance between the substrate
and the upper prism facet (diamonds) and the observed half-width of the
column along the substrate (circles). The inset diagram in the upper left
shows the approximate cross section of the crystal at $t=41$ seconds
(although the detailed structure of the columnar walls was not determined;
see Figure \protect\ref{sample}). Lines show model crystal calculations
using $(A,\protect\sigma _{0})_{basal}=(1,0.025\%),$ $(A,\protect\sigma %
_{0})_{prism}=(0.15,0.17\%),$ $(R,H)_{initial}=(11.1,15.2),$ and
supersaturations $\protect\sigma _{\infty }=$ 1.6, 2.4, and 3.6 percent, as
labeled. The images below the graph show calculated crystal cross sections
at $t=45$ (for the 1.6\% and 2.4\% models) or $t=29$ (for the 3.6\% model).}
  \label{xtal3sinf}
\end{figure}

\begin{figure}[tbp] 
  \centering
  \includegraphics[width=5.67in,height=5.23in,keepaspectratio]{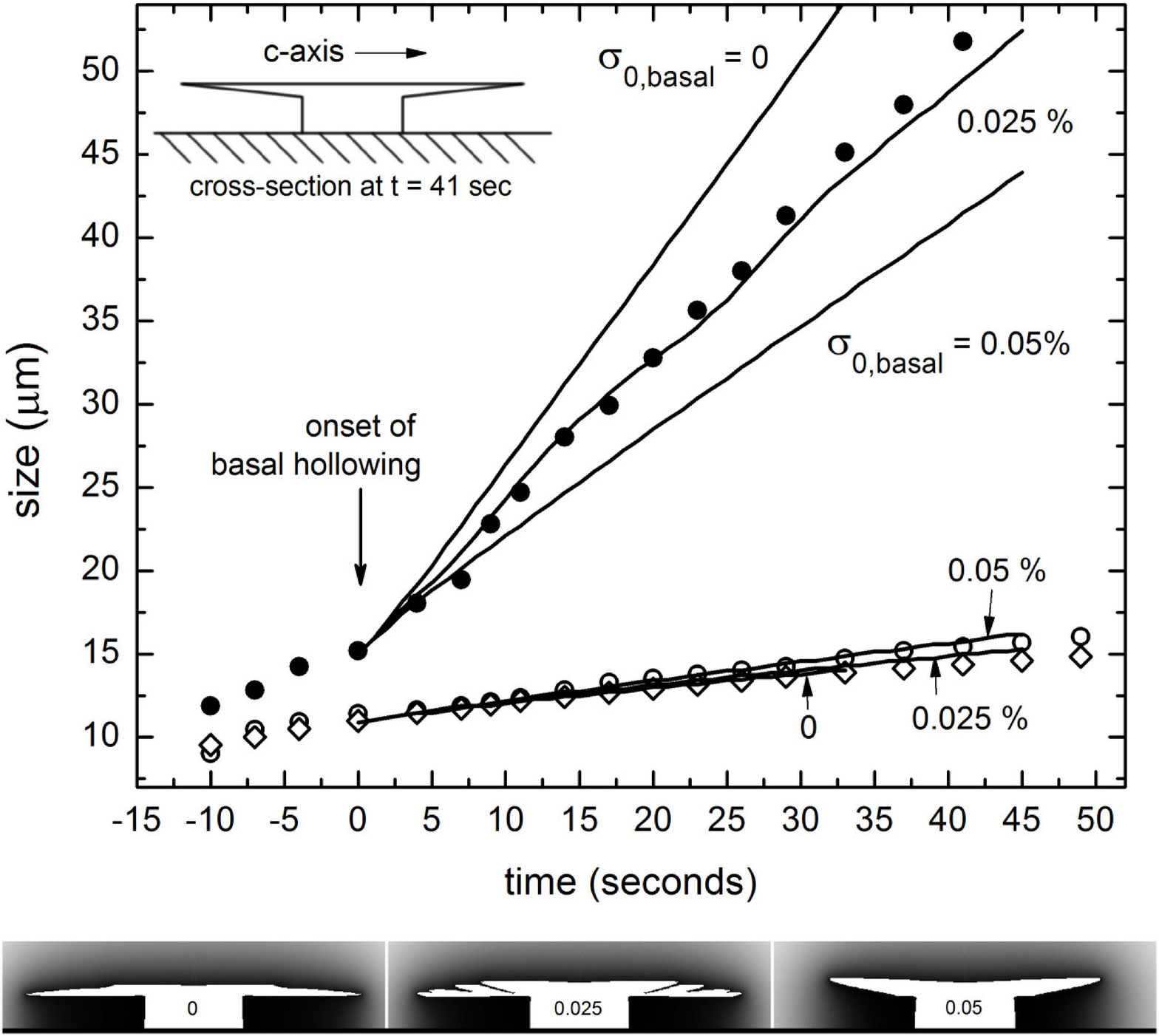}
  \caption{The same experimental data as in
Figure \protect\ref{xtal3sinf}, but plotted with different theoretical
models. Lines show model crystal calculations using $(A,\protect\sigma %
_{0})_{prism}=(0.15,0.17\%),$ $(R,H)_{initial}=(10.2,13.3),$ $A_{basal}=1,$ $%
\protect\sigma _{\infty }=$ 2.4 percent, with $\protect\sigma _{0,basal}=$
0, 0.025, and 0.05 percent, as labeled. The images below the graph again
show calculated crystal cross sections as in Figure \protect\ref{xtal3sinf}.}
  \label{xtal3s0}
\end{figure}

Figures \ref{xtal3sinf} and \ref{xtal3s0} show data and models for a crystal
grown at an experimentally set $\sigma _{\infty }=3.9$ percent, high enough
to produce strong hollowing of the basal facets at this temperature. In this
case the crystal morphology was similar to that seen in the final stages of
growth in Figure \ref{sample}, showing deep basal hollowing with a
thin-walled hollow columnar morphology. Although complex in its fine
details, the morphology was essentially that of half a hollow column, so to
an adequate approximation we were able to numerically model the crystal
using our 2D cylindrically symmetrical cellular automata code. As before, we
set $\alpha _{prism}(\sigma _{surf})$ to the intrinsic value, and adjusted $%
\sigma _{\infty }$ and $\alpha _{basal}$ to fit the data.

Figure \ref{xtal3sinf} shows three models in which we fixed $\alpha _{basal}$
and varied $\sigma _{\infty },$ centered about our best-fit model. Since it
took some time for the supersaturation to stabilize in the experiment, we
began the models when the basal hollowing was first observed in this
crystal, shown as $t=0$ in the figures. In the models, the supersaturation
field was allowed to fully relax before commencing crystal growth, and in
all cases basal hollowing appeared very quickly. In Figure \ref{xtal3sinf}
we again see that higher $\sigma _{\infty }$ values yielded faster growth
rates for both facets, as expected. And our best-fit $\sigma _{\infty }=2.4$
percent was lower than the $\sigma _{\infty }$ = 3.9 percent set in the
experiment, as expected.

Figure \ref{xtal3s0} shows the same crystal data along with three models in
which we fixed $\sigma _{\infty }$ and varied $\sigma _{0,basal},$ again
centered about our best-fit model. And, as with the previous crystal, we see
that reducing the nucleation barrier on the basal facet increased the basal
growth while slightly reducing the prism growth rate. By adjusting both $%
\sigma _{\infty }$ and $\sigma _{0,basal},$ we produced our best-fit model
with $\sigma _{\infty }=2.4$ percent and $\sigma _{0,basal}=0.025$ percent.
Note also that a morphological transition appeared in the models as we
changed $\sigma _{0,basal}.$ With $\sigma _{0,basal}=0$ the prism facets
showed convex shapes, while at $\sigma _{0,basal}=0.05$ percent the prism
facets were concave. Although this transition may become altered with full
3D modeling, we suggest that it may be a robust feature in diffusion-limited
faceted crystal growth.

Our overall conclusions from this high-$\sigma _{\infty }$ crystal are quite
different from the previous low-$\sigma _{\infty }$ crystal. For the high-$%
\sigma _{\infty }$ data shown in Figures \ref{xtal3sinf} and \ref{xtal3s0},
the crystal growth rates and morphology cannot be adequately modeled using
the intrinsic attachment coefficients. In particular, modeling the high-$%
\sigma _{\infty }$ behavior required $\sigma _{0,basal}\approx 0.025$
percent, over a factor of 10 smaller than the low-$\sigma _{\infty }$
crystal, and approximately a factor of 30 smaller than the intrinsic $\sigma
_{0,basal}.$ This discrepancy is simply too large to be the result of
systematic effects in the experiment or the modeling, so some other physical
mechanism is necessary to explain the observations.

The SDAK instability described in \cite{standardmodel} provides a natural
explanation for both the low-$\sigma _{\infty }$ and high-$\sigma _{\infty }$
data presented here. At low $\sigma _{\infty },$ the facets are large and
flat, so the intrinsic $\alpha _{basal}$ and $\alpha _{prism}$ can
adequately describe the growth behavior. At high $\sigma _{\infty },$
however, the SDAK instability on the basal facets produces a thin edge with
a much reduced nucleation barrier, thus resulting in a thin-walled hollow
columnar morphology. Including the SDAK effect allows us to qualitatively
explain the morphologies and quantitatively fit the measured growth rates.

\section{Conclusions}

In summary, we have examined the growth of ice crystals from water vapor at
a temperature of -5.15 C, in an atmosphere of air at one bar. Although
detailed data from only two test crystals are presented above, observations
of additional crystals indicated that our overall conclusions are robust
from sample to sample. Strong basal hollowing was seen in all high-$\sigma
_{\infty }$ samples, while all low-$\sigma _{\infty }$ grew as simple
prisms. In addition, while complicating effects arising from substrate
interactions, modeling systematics, crystal-to-crystal variations,
uncertainties in determining the supersaturation accurately, etc., were not
negligible, we believe that these effects do not substantially affect our
overall conclusions.

Our main conclusion is that our observations are consistent with the model
presented in \cite{standardmodel}. Our assumptions include: 1) the intrinsic 
$\alpha _{surf}(\sigma _{surf},T)$ are given by the measurements presented
in \cite{prisms}, 2) the $\alpha _{surf}(\sigma _{surf},T)$ are not
substantially altered by a background air pressure of one bar, and 3) the
numerical modeling method described in \cite{kglca} is adequate. Given these
assumptions, the measurements presented above then strongly support our
hypothesis that SDAK effects are present on the basal facet at -5.15 C. 

We believe that the comprehensive model of ice crystal growth presented in 
\cite{standardmodel} is at least on the right track. After conceiving it,
the model immediately made a clear prediction that SDAK effects should be
present on the basal facet near -5 C. When we subsequently performed the
above experiments to look for these effects, they were clearly present,
essentially just as model predicted. Our numerical modeling of the data
indicates that the rapid basal growth associated with basal hollowing is
consistent with an SDAK instability, and this behavior is difficult to
explain otherwise.

Additional precision ice crystal growth measurements at other temperatures,
together with additional modeling, should further elucidate the underlying
molecular dynamics governing ice growth behavior. From this we hope to
better understand the ice surface structure and dynamics, and how these
change with temperature and other factors on the principal facets. And by
using ice as a case study, we hope to gain additional insights into the
many-body surface physics that governs crystal growth more generally.

\end{document}